# Uniform and Accelerated Circular Motion with a Smartphone: A No-Code, AI-Generated Browser Laboratory

**Josep Ll. Suñer, Francisco M. Muñoz-Pérez, Juan C. Castro-Palacio, and Juan A. Monsoriu,** *Universitat Politècnica de València, Valencia, Spain*

**Martín Monteiro,** *Universidad ORT Uruguay, Montevideo, Uruguay*

**Cecilia Stari and Arturo C. Martí,** *Universidad de la República, Montevideo, Uruguay*

Smartphones have become a standard measurement instrument in the physics laboratory.[1,2] Their built-in accelerometers, gyroscopes, magnetometers, and cameras have been used to investigate a wide range of phenomena in mechanics, and rotational motion in particular has proven especially well-suited to smartphone-based experiments.[3–9] A recurring limitation, however, is that most experiments rely on precompiled sensor apps whose interfaces cannot be tailored to a specific activity, and until recently creating customized smartphone laboratories required programming knowledge beyond what most physics teachers possess.

This paper makes two connected contributions. First, we present a validated smartphone experiment for uniform circular motion (UCM) and uniformly accelerated circular motion (UACM), in which the angular velocity and the angular acceleration are obtained from two independent smartphone channels and checked against independent video analysis with Tracker.[10,14] Second, we show that generative AI can serve as a no-code development tool for the physics laboratory: the browser-based application used here was produced entirely through natural-language prompting, with no manual coding, extending to rotational motion the approach we recently introduced for acoustic experiments.[11] The emphasis is therefore not only on what is measured, but also on how the measuring instrument itself was built.

## The experiment

The experimental setup (Fig. 1) consists of a rotating platform on which the smartphone is placed, a low-friction vertical shaft, and a drum around which a string is wound; the string passes over a pulley and is attached to a hanging mass. The smartphone simultaneously records the orientation angles and the components of the angular velocity by means of its built-in sensors, while the platform rotates freely about its axis.

The main advantage of this arrangement is that the same device makes it possible to study two types of motion simply by changing how it is set into operation. To study UCM, the hanging mass is removed and a small initial angular velocity is imparted to the platform by hand. Owing to the low friction of the system, the angular velocity remains practically constant for several seconds. To study UACM, the hanging mass is attached and released. The string tension exerts an approximately constant torque on the drum, producing a nearly constant angular acceleration. In both cases the smartphone records the time evolution of the rotation angle and of the angular velocity, allowing the kinematic equations of rotation to be verified experimentally.

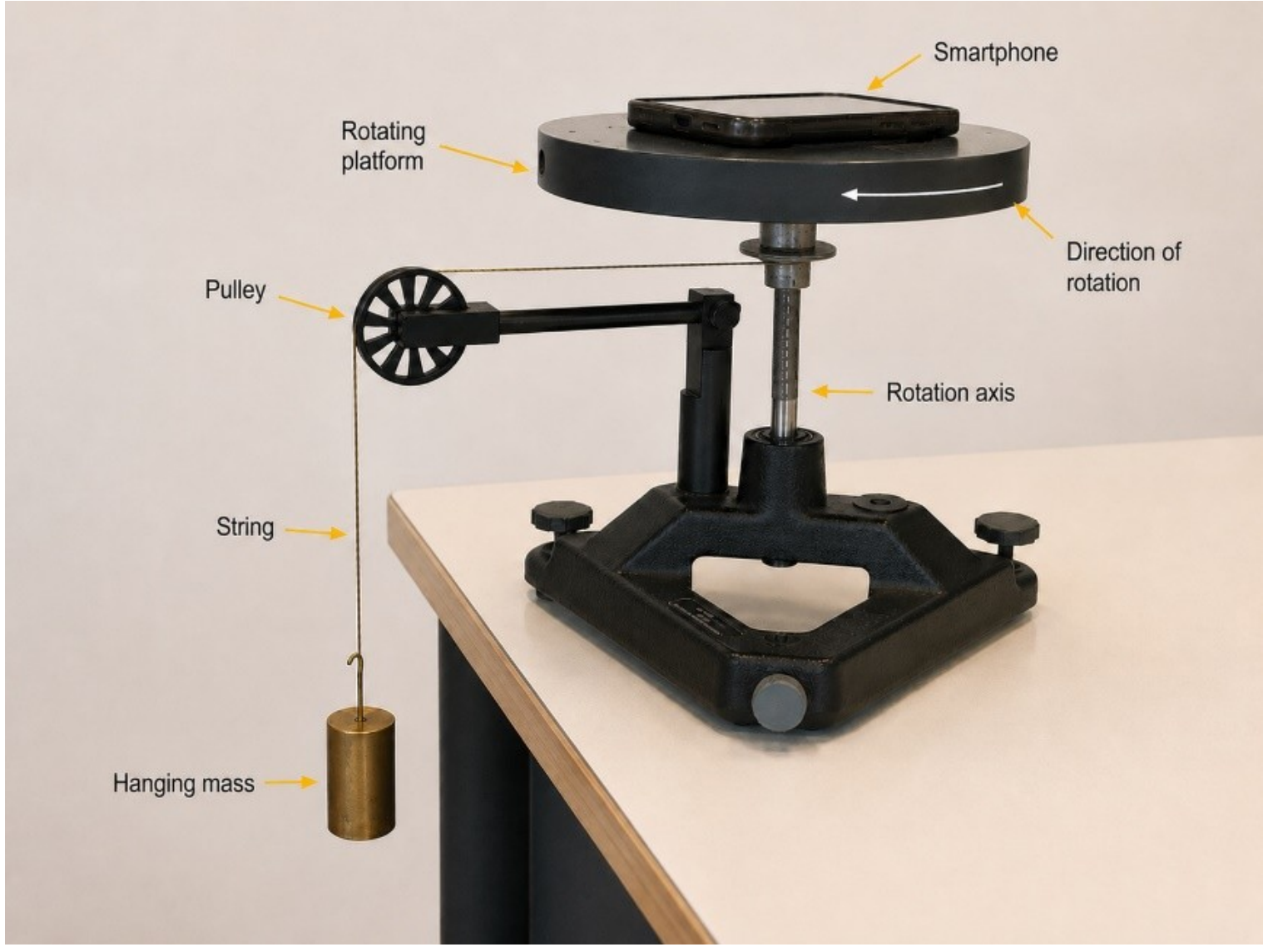

**Fig. 1.** Experimental setup: a smartphone on a rotating platform mounted on a low-friction vertical shaft. A string wound around the drum passes over a pulley and is attached to a hanging mass, which drives the uniformly accelerated motion.

## The AI-generated laboratory

The measurement application, *SmartPhysics: Rotation Lab*,[15] is a single self-contained HTML file that requires no installation. It was developed in one conversational session with Claude Sonnet 4.6 (Anthropic), following the workflow described in Ref. 11, by refining the specification until the interface met our requirements. Once the application had reached its final form, the assistant was asked to produce a single consolidated prompt reproducing it. It is this consolidated prompt, rather than the initial one, that we reproduce below and provide in full as supplementary material, because it regenerates the final version in one step: starting from it, a single prompting iteration is sufficient, with no correction of the generated code. At no stage was any line of code written or edited by hand.

The consolidated prompt begins by fixing the scope and the delivery format of the tool:

> "Create a single-file mobile-first HTML physics lab tool called 'SmartPhysics: Rotation Lab' that reads the device's gyroscope and orientation sensors and plots the data live. The entire app should be one self-contained .html file with no external dependencies except Google Fonts."

It continues with separate blocks specifying the visual layout, the quantities to be plotted and their units, the sensor-access logic, and the export options. Only the block describing the measured quantities is specific to the physics of this activity, which is what makes the prompt reusable as a template. The complete prompt and the generated HTML file are provided as supplementary material.

The application accesses both the DeviceOrientation API and the raw gyroscope data through standard browser Web APIs. It is important to emphasize that the orientation sensor is not a separate physical hardware component; rather, the browser's orientation angles are generated via internal sensor fusion, integrating the physical gyroscope's rate measurements with accelerometer data. Reading both APIs therefore allows us to test the internal consistency between the device's integrated software output and its raw sensor readings, while video analysis provides the only fully independent physical validation.

Because the application relies on the DeviceOrientationEvent and DeviceMotionEvent interfaces, it requires a mobile browser that exposes these APIs and a page served over HTTPS; on iOS, sensor access must additionally be granted through an explicit user gesture, which the application requests on the first tap of the START button. Desktop browsers, and mobile browsers that do not expose the motion sensors, load the interface but record no data. The measurements reported here were obtained with a Redmi Note 13 Pro 5G smartphone (Xiaomi HyperOS 2.0.206.0.VNREUXM) using Google Chrome.

The sampling rate and buffer length are adjustable, the plots support zoom and pan, and the recorded data can be exported as CSV files and screen captures as PNG images for quantitative analysis with standard software. In the experiments reported here the sampling rate was set to 50 Hz, and the effective rate delivered by the device matched this nominal value.

It should be stressed that, in the activity as implemented here, the AI assistant intervenes before instruction rather than during it: its role is AI-assisted tool creation, and students interact only with the resulting application, not with the assistant itself. This is a deliberate choice, since it keeps the laboratory session focused on the physics while still giving the instructor full control over the measurement interface. A natural extension, which we have not yet evaluated in the classroom, is to have students write and refine the prompt themselves: deciding which quantities a measurement application should display requires them to identify in advance the relevant variables, the expected functional dependences, and what would constitute a meaningful comparison between measurement channels.[11–13]

Figure 2 shows two representative screenshots of the laboratory during the experiments. In both cases, the upper panel displays the orientation angles and the lower panel the three components of the angular velocity. Since the only appreciable rotation is about the vertical axis, only the azimuth angle and the corresponding angular velocity component take significant values, while the remaining series stay close to zero. For UCM [Fig. 2(a)] the angular velocity is approximately constant and the angle grows linearly in time, whereas for UACM [Fig. 2(b)] the angular velocity increases linearly and the angle exhibits the characteristic quadratic evolution of motion with constant angular acceleration.

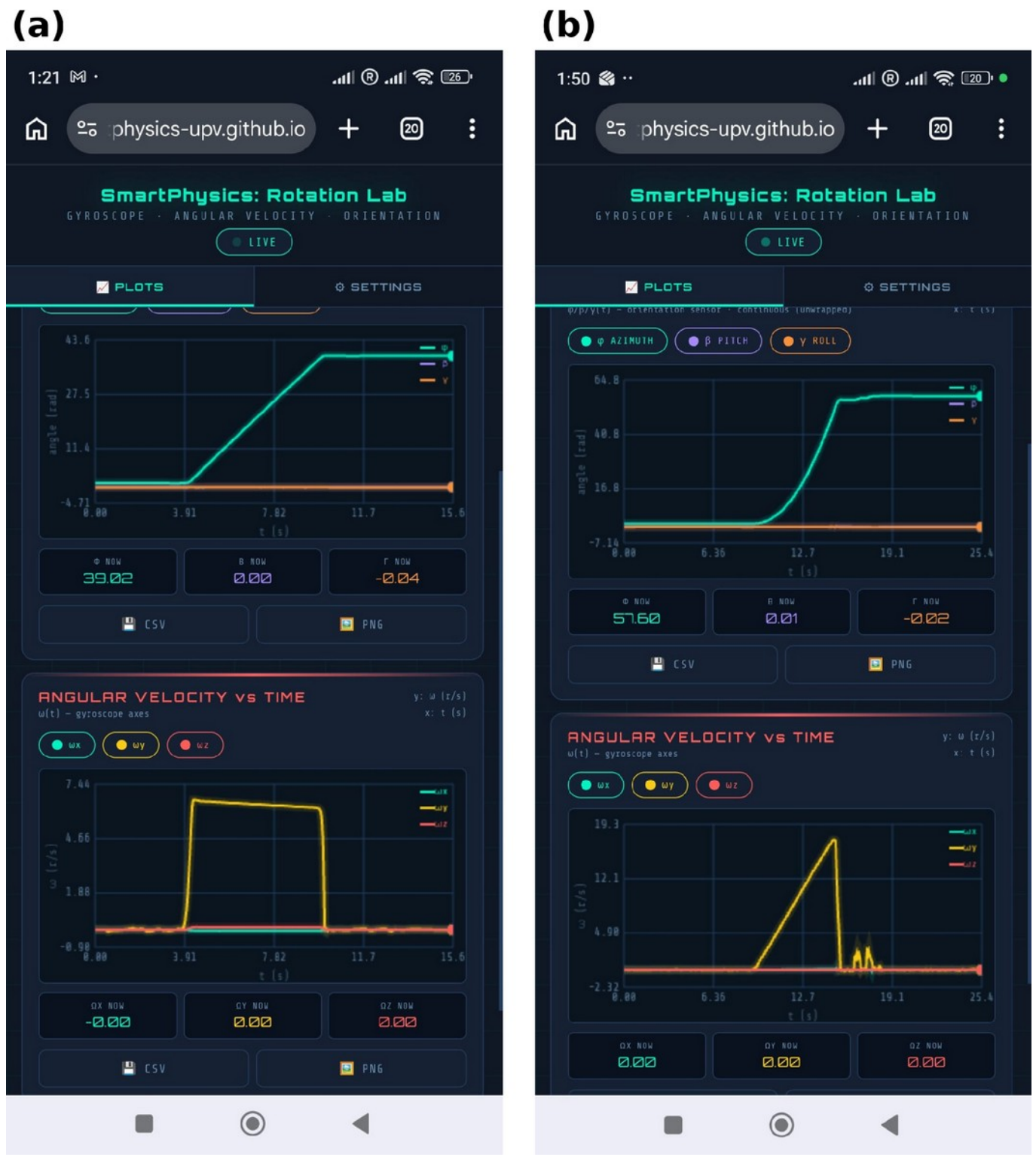

**Fig. 2.** Screenshots of the AI-generated application during (a) uniform circular motion and (b) uniformly accelerated circular motion. Upper panels: orientation angles (azimuth, pitch, roll). Lower panels: components of the angular velocity measured by the gyroscope.

## Results: uniform circular motion

Figure 3(a) shows the angular position obtained from the orientation API over an analyzed interval of 4.1 s, together with the linear fit $\theta(t) = \omega t + \theta_0$. The slope directly yields the angular velocity, $\omega = 6.110(5)$ rad/s, with a coefficient of determination $R^2 = 0.9999$, confirming the excellent linear behavior expected for UCM. Figure 3(b) shows the angular velocity recorded simultaneously by the gyroscope; here no fit is needed, and the mean value is $\omega = 6.10(8)$ rad/s. The slopes and their uncertainties were obtained with standard spreadsheet functions (LINEST for the regressions, and AVERAGE and STDEV for the mean values); the figure in parentheses indicates the uncertainty in the last digit. The two values differ by only 0.1%, well within the quoted uncertainties, confirming the internal consistency of the browser's sensor-fusion pipeline. As a fully independent external check, video analysis of the motion with Tracker[14] gives $\omega = 6.13(2)$ rad/s ($R^2 = 0.9986$), in agreement with both smartphone measurements to better than 0.5% (Table I).

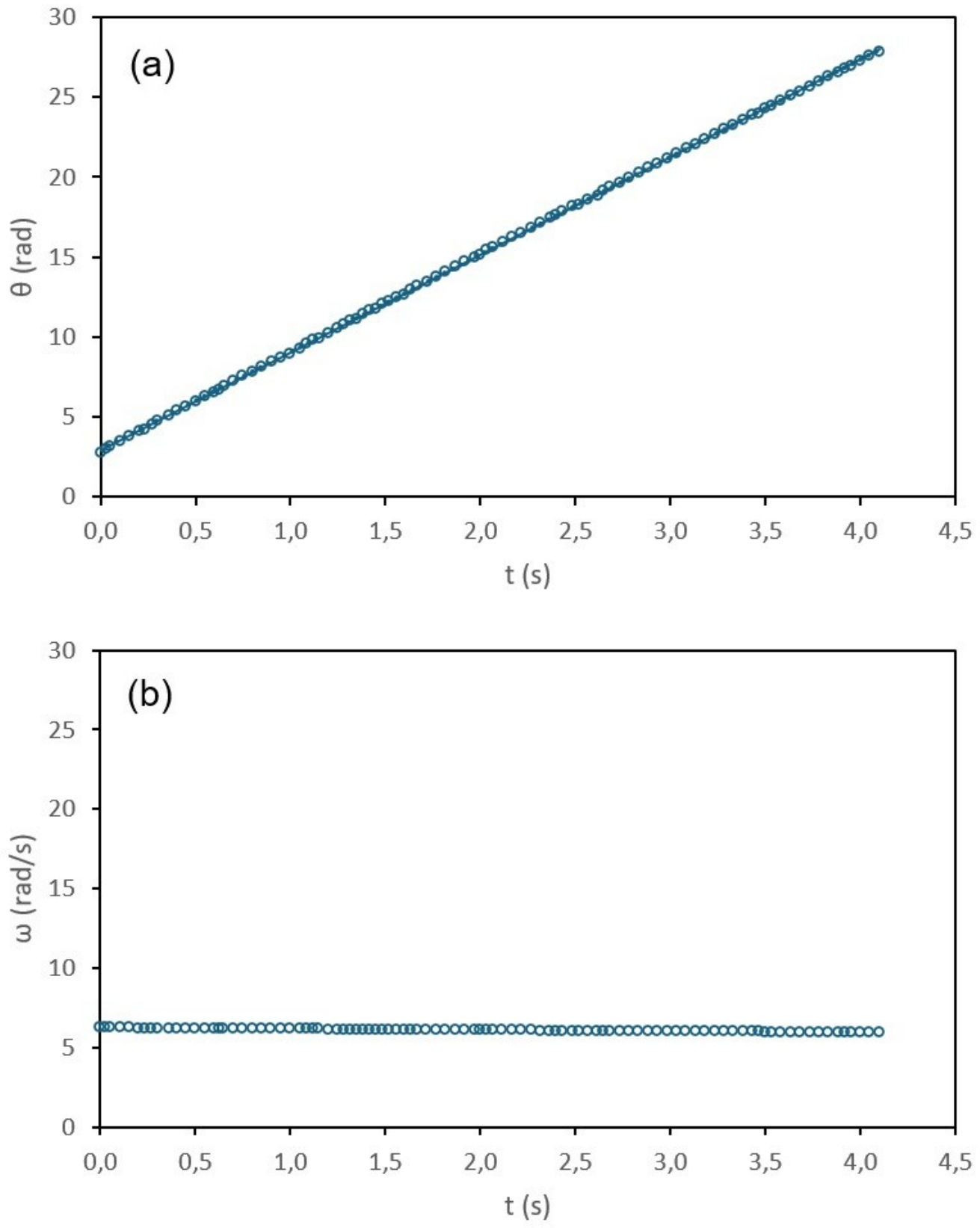

**Fig. 3.** Uniform circular motion. (a) Angular position from the orientation API and linear fit; the slope gives $\omega = 6.110(5)$ rad/s ($R^2 = 0.9999$). (b) Angular velocity from the gyroscope; the mean value is $\omega = 6.10(8)$ rad/s.

## Results: uniformly accelerated circular motion

With the hanging mass attached, the angular position follows the quadratic dependence expected for constant angular acceleration. Fitting the orientation API data over an analyzed interval of 2.7 s to $\theta(t) = \frac{1}{2}\alpha t^2 + \omega_0 t + \theta_0$ [Fig. 4(a)] yields $\alpha = 7.58(2)$ rad/s$^2$ with $R^2 \approx 1$. The gyroscope data [Fig. 4(b)] increase linearly in time, as predicted; the linear fit $\omega(t) = \alpha t + \omega_0$ gives $\alpha = 7.614(8)$ rad/s$^2$ ($R^2 = 0.9999$). The difference between the two smartphone determinations is only 0.5%. Video analysis with Tracker, fitting the angular position to the same quadratic model, gives $\alpha = 7.56(2)$ rad/s$^2$ ($R^2 = 0.9998$). The maximum discrepancy among the three methods is below 1% (Table I), a solid experimental validation of the laboratory. The videos used for the Tracker analysis of both motions are provided as supplementary material.

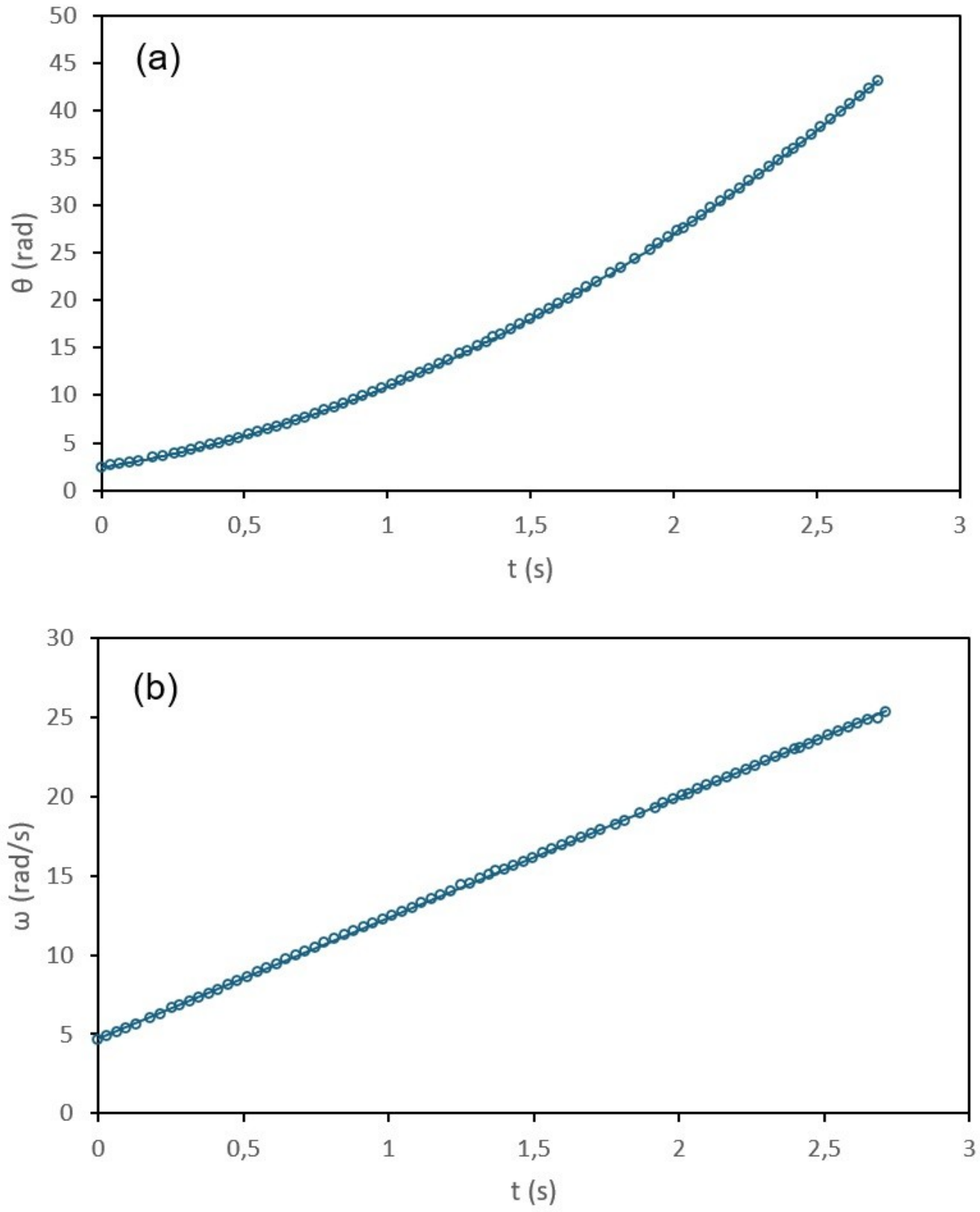

**Fig. 4.** Uniformly accelerated circular motion. (a) Angular position from the orientation API and quadratic fit, giving $\alpha = 7.58(2)$ rad/s$^2$. (b) Angular velocity from the gyroscope and linear fit, giving $\alpha = 7.614(8)$ rad/s$^2$.

**Table I.** Angular velocity of the UCM and angular acceleration of the UACM obtained with the two smartphone measurement channels and with video analysis (Tracker). Since the orientation API and the gyroscope share the same underlying hardware, linked via sensor fusion, the video analysis provides the primary independent validation.

| **Method** | **ω (rad/s), UCM** | **α (rad/s²), UACM** |
|---|---|---|
| Orientation API | 6.110(5) | 7.58(2) |
| Gyroscope | 6.10(8) | 7.614(8) |
| Video analysis (Tracker) | 6.13(2) | 7.56(2) |

## Conclusions

The same simple setup allows both uniform and uniformly accelerated circular motion to be studied using only a smartphone as the measuring instrument. In the first case, a constant angular velocity is verified to produce a linear evolution of the angular position; in the second, an approximately constant angular acceleration is shown to produce, simultaneously, a linear evolution of the angular velocity and a quadratic evolution of the angular position. The results obtained from the orientation API (sensor fusion) and the raw gyroscope data show internal agreement below 0.5%, and both align with the independent video-analysis technique to better than 1%. This multi-channel approach allows students not only to verify the fundamental kinematic equations of rotation, but also to appreciate how mobile Web APIs process raw inertial measurements.

Beyond kinematic verification, the pedagogical value of this activity lies in software customization. Instructors can eliminate distracting features common in commercial software (such as complex calibration menus or unneeded sensor channels) and focus student attention exclusively on the relevant variables ($\theta$ and $\omega$). Furthermore, the setup is robust against minor experimental imperfections: because angular velocity measurements from gyroscopes depend on the rotation rate around the z-axis rather than spatial position, slight off-center placement of the smartphone on the rotating platform does not introduce systematic errors in $\omega$ or $\alpha$.

More important is the way the laboratory itself was developed. As in our previous work,[11] the application was generated entirely by artificial intelligence from natural-language instructions, without developing a dedicated mobile app. Instead of adapting classroom activities to existing software, instructors can now design the software around their own pedagogical objectives (presenting only the variables relevant to the activity, simplifying the interface, and removing unnecessary controls) using AI as a programming tool, which drastically reduces development time and eliminates the need for advanced programming skills.[11–13] In this approach, smartphones cease to be mere data-acquisition devices and become complete experimental platforms that record, display, and export results in real time. The combination of built-in sensors, AI-generated web applications, and simple analysis techniques makes it possible to develop low-cost, easily reproducible physics experiments with accuracy sufficient to test quantitatively the theoretical models studied in introductory physics courses.

**Acknowledgments**

This work was funded by the Polytechnic University of Valencia, Spain (grants PIME/23-24/374, PIME/25-26/578, and EICE SmartSTEM).

**Supplementary material**

The supplementary material includes the complete natural-language prompt used to generate the application, the generated HTML file, and the videos of the UCM and UACM experiments used for the Tracker analysis.

## References

1. M. Monteiro and A. C. Martí, "Resource Letter MDS-1: Mobile devices and sensors for physics teaching," Am. J. Phys. **90**, 328–343 (2022).
2. J. Kuhn and P. Vogt, *Smartphones as Mobile Minilabs in Physics* (Springer, Heidelberg, 2022).
3. M. Monteiro, C. Cabeza, A. C. Martí, P. Vogt, and J. Kuhn, "Angular velocity and centripetal acceleration relationship," Phys. Teach. **52**, 312–313 (2014).
4. J. C. Castro-Palacio, L. Velázquez, J. A. Gómez-Tejedor, F. J. Manjón, and J. A. Monsoriu, "Using a smartphone acceleration sensor to study uniform and uniformly accelerated circular motions," Rev. Bras. Ensino Fís. **36**(2), 2315 (2014).
5. R. Pörn and M. Braskén, "Interactive modeling activities in the classroom—rotational motion and smartphone gyroscopes," Phys. Educ. **51**, 065022 (2016).
6. V. L. B. de Jesus, C. A. C. Pérez, A. L. de Oliveira, and D. G. G. Sasaki, "Understanding the gyroscope sensor: A quick guide to teaching rotation movements using a smartphone," Phys. Educ. **54**, 015003 (2018).
7. I. Salinas, M. H. Giménez, J. A. Monsoriu, and J. A. Sans, "Demonstration of the parallel axis theorem through a smartphone," Phys. Teach. **57**, 340–341 (2019).
8. I. Salinas, M. Monteiro, A. C. Martí, and J. A. Monsoriu, "Analyzing the dynamics of a yo-yo using a smartphone gyroscope sensor," Phys. Teach. **58**, 569–571 (2020).
9. A. C. Martí, M. Monteiro, and C. Stari, "The circular Atwood machine," Phys. Teach. **62**, 150–151 (2024).
10. M. Monteiro, C. Cabeza, C. Stari, and A. C. Martí, "Smartphone sensors and video analysis: Two allies in the physics laboratory battle field," Phys. Educ. **55**, 035012 (2020).
11. J. Ll. Suñer, F. M. Muñoz-Pérez, P. Yuste, J. C. Castro-Palacio, and J. A. Monsoriu, "Generating interactive smartphone physics laboratories using artificial intelligence," Phys. Educ. **61**, 035041 (2026).
12. Y. Ben-Zion, R. Einhorn Zarzecki, J. Glazer, and N. D. Finkelstein, "Leveraging AI for rapid generation of physics simulations in education: Building your own virtual lab," Phys. Teach. **63**, 424–427 (2025).
13. D. Kim, T. Havarko, S. Küchemann, J. Kuhn, P. Vogt, and D. Rakestraw, "Star trail analysis with an AI coinvestigator," Phys. Teach. **64**, 330–331 (2026).
14. D. Brown, W. Christian, and R. M. Hanson, *Tracker Video Analysis and Modeling Tool*, https://physlets.org/tracker/.
15. SmartPhysics, "Rotation Lab" (2026), https://smartphysics.webs.upv.es/rotation-lab.